\pdfoutput=1
\documentclass[reprint,aps,prc,floatfix,nofootinbib,amsmath,amssymb,superscriptaddress]{revtex4-2}

\usepackage{graphicx}
\usepackage{bm}
\usepackage[colorlinks=true,linkcolor=blue,citecolor=blue,urlcolor=blue]{hyperref}

\newcommand{\sNN}{\sqrt{s_{NN}}}
\newcommand{\meanpT}{\langle\!\langle p_T \rangle\!\rangle}
\newcommand{\correl}{\langle \Delta p_{T,i}\Delta p_{T,j} \rangle}
\newcommand{\Tr}{\operatorname{Tr}}
\newcommand{\Var}{\operatorname{Var}}
\newcommand{\Cov}{\operatorname{Cov}}

\begin{document}

\title{Equilibrium baseline for transverse-momentum correlations in Au+Au collisions at \texorpdfstring{$\sqrt{s_{NN}} = 3$--$39$~GeV}{sqrt(sNN) = 3-39 GeV}: Conservation laws, resonance decays, quantum statistics, and collective flow}

\author{Oleh Savchuk}
\email{savchuk@frib.msu.edu}
\affiliation{International Research Laboratory on Nuclear Physics and Astrophysics, Michigan State University and CNRS, East Lansing, Michigan 48824, USA}

\date{\today}

\begin{abstract}
Event-by-event transverse-momentum correlations measured by STAR in Au+Au collisions exhibit a nonmonotonic energy dependence with a minimum near $\sNN \approx 5$--$8$~GeV and a steep rise toward $3$~GeV. Before interpreting these features in terms of a softened equation of state or a QCD critical point, the baseline from an equilibrated hadron resonance gas must be established. Here we compute the two-particle correlator $\correl$ in thermal and chemical equilibrium from $\sNN = 3$ to $39$~GeV for 0--5\% and 30--40\% collisions, incorporating resonance decays, Bose-Einstein statistics, Bjorken longitudinal and transverse radial flow, and global conservation of energy, baryon number, electric charge, strangeness, and transverse momentum. Imposing conservation laws via a saddle-point expansion of the microcanonical ensemble yields the variance of any additive observable as a Schur complement of grand-canonical covariances. Although the equilibrium gas undergoes the transition from baryon- to meson-dominated matter, composition fluctuations cancel the large negative ``mixing'' term that arises when species multiplicities are fixed event by event, leaving only a much smaller negative conservation residual. Across $\sNN = 11.5$--$39$~GeV, the equilibrium baseline with strong flow underpredicts the data by a factor of $3$--$4$ in 0--5\% and $\approx 6.5$ in 30--40\% collisions; at fixed-target energies ($3$--$5.2$~GeV), where the data are largest and positive, the baseline turns negative. Accounting for the data via radial-flow fluctuations at fixed energy requires $\sigma_\beta/\beta_s \approx 6$--$10\%$ in central and $\approx 22\%$ in 30--40\% collider-energy collisions, growing faster with decreasing centrality than independent-source scaling.
\end{abstract}

\maketitle

\section{Introduction}
\label{sec:intro}

Event-by-event fluctuations of the mean transverse momentum were among the earliest observables proposed to probe thermalization and critical phenomena in relativistic heavy-ion collisions~\cite{Voloshin1999}. Experimentally, the STAR Collaboration characterizes these fluctuations through the two-particle transverse-momentum correlator~\cite{STAR2019,STAR2026}
\begin{align}
\correl &= \frac{1}{N_{\mathrm{ev}}} \sum_{k=1}^{N_{\mathrm{ev}}} \frac{C_k}{N_k(N_k - 1)}, \nonumber\\
C_k &= \sum_{i=1}^{N_k} \sum_{\substack{j=1 \\ j \neq i}}^{N_k} \bigl(p_{T,i} - \meanpT\bigr)\bigl(p_{T,j} - \meanpT\bigr),
\label{eq:correlator}
\end{align}
where $N_k$ is the number of accepted charged hadrons in event $k$, $\meanpT$ is the inclusive event-averaged transverse momentum, and $C_{p_T} = \sqrt{\correl}/\meanpT$ measures the relative dynamical correlation. During the first phase of the RHIC Beam Energy Scan (BES-I), STAR found that $C_{p_T}$ in central Au+Au collisions increases smoothly with $\sNN$ from $7.7$ to $200$~GeV~\cite{STAR2019}. Extending these measurements down to $\sNN = 3$~GeV in the BES-II fixed-target (FXT) program revealed a striking nonmonotonic trend in central collisions, with a pronounced minimum around $\sNN \approx 5$--$8$~GeV followed by a sharp rise toward lower energies~\cite{STAR2026}---a behavior noted for its potential sensitivity to the high-density equation of state and the QCD critical point.

Several dynamical mechanisms can generate nonzero $\correl$: event-by-event fluctuations of the initial transverse size converted into radial-flow fluctuations by hydrodynamic expansion~\cite{Bozek2012,Schenke2020}, incomplete thermalization and local momentum conservation~\cite{Gavin2004}, hadronic rescattering modeled in JAM2~\cite{Liu2026}, and changes in hadronic composition with collision energy. In particular, Reichert, Steinheimer, and Bleicher~\cite{Reichert2026} argued that a minimum near $\sNN \approx 5$--$8$~GeV need not signal critical behavior: in a two-component mixture of pions and protons with distinct mean transverse momenta, fixing particle numbers produces a negative contribution proportional to $(\langle p_T\rangle_p - \langle p_T\rangle_\pi)^2$ whose magnitude peaks where protons and pions are equally abundant, right across the baryon-to-meson transition.

Before attributing the measured correlations to any dynamical or critical mechanism, however, one must quantify the equilibrium baseline arising from standard statistical and kinematic correlations---global conservation laws, resonance decays, and quantum statistics---in an expanding hadron gas. While an unconstrained gas of independently emitted Boltzmann particles has $\correl = 0$ identically, each of these baseline effects induces correlations of order $1/N$. Whether their combined action in a realistic hadron resonance gas (HRG) can account for a significant fraction of the measured signal, or generate the low-energy FXT rise, has not previously been established. Here we compute this complete equilibrium baseline across $\sNN = 3$--$39$~GeV for 0--5\% and 30--40\% central Au+Au collisions.

To enforce global conservation laws in thermal equilibrium, we pass from the grand-canonical ensemble (GCE) to the microcanonical ensemble (MCE) via Laplace and Fourier projections of the partition function. In the thermodynamic limit, this saddle-point expansion yields a multivariate Gaussian structure that is well known for multiplicity fluctuations~\cite{Becattini2005,Begun2006,Hauer2008} and conserved-charge cumulants~\cite{Vovchenko2020}. Extending the same framework to momentum-weighted observables---where energy conservation plays a central role---expresses the microcanonical variance concisely as a Schur complement~\cite{Schur1917,Banachiewicz1937,Haynsworth1968,Zhang2005} of GCE covariances. Specifically, this work provides (i)~a closed-form conservation baseline for $\correl$ incorporating energy, baryon number, electric charge, strangeness, and transverse momentum, (ii)~a quantitative evaluation using the full PDG2025 hadron spectrum with resonance decays, Bose-Einstein statistics, and collective longitudinal and radial flow compared against STAR BES-I, BES-II collider, and FXT data, and (iii)~a direct test of the baryon--meson mixing mechanism within a self-consistent equilibrium ensemble.

\section{Conservation laws from the saddle point}
\label{sec:saddle}

\subsection{Microcanonical ensemble as a conditioned grand-canonical ensemble}
\label{sec:mce}

Let $\hat{\bm{Q}} = (\hat{E}, \hat{B}, \hat{Q}, \hat{S})$ denote the globally conserved four-vector of energy, baryon number, electric charge, and strangeness. We obtain the MCE at fixed $\bm{Q}_0$ from the GCE by writing the energy $\delta$ function as an inverse Laplace transform (a Bromwich integral in $\beta$) and the Kronecker $\delta$ functions for the integer charges as Fourier integrals over conjugate phases. Introducing $\bm{\tau} = (-\beta, i\phi_B, i\phi_Q, i\phi_S)$, the microcanonical expectation value of an observable $F$ takes the form
\begin{equation}
\langle F \rangle_{\mathrm{MCE}} = \frac{\int d\bm{\tau}\, e^{-\bm{\tau}\cdot\bm{Q}_0} \Tr\bigl[F\, e^{\bm{\tau}\cdot\hat{\bm{Q}}}\bigr]}{\int d\bm{\tau}\, e^{-\bm{\tau}\cdot\bm{Q}_0} \Tr e^{\bm{\tau}\cdot\hat{\bm{Q}}}}.
\label{eq:mce_avg}
\end{equation}
To extract the variance of an additive one-body observable $X$, we set $F = e^{sX}$ and introduce the extensive GCE cumulant generating function $\Phi(s, \bm{\tau}) = \ln \Tr e^{sX + \bm{\tau}\cdot\hat{\bm{Q}}}$. At $s=0$, the $\bm{\tau}$ integral is dominated by the saddle point $\bm{\tau}^*$ satisfying $\langle \hat{\bm{Q}} \rangle_{\mathrm{GCE}}(\bm{\tau}^*) = \bm{Q}_0$. This condition defines the temperature and chemical potentials of the equivalent GCE (the chemical freeze-out parameters, with $\mu_Q$ and $\mu_S$ fixed by the colliding-system ratios $Q/B = 79/197$ and $S = 0$).

Along the integration contour, fluctuations around the saddle point are purely imaginary, $\bm{\tau} = \bm{\tau}^* + i\bm{\omega}$ with real $\bm{\omega}$. Expanding $\Phi(s, \bm{\tau}) - \bm{\tau}\cdot\bm{Q}_0$ to second order in $s$ and $\bm{\omega}$ and performing the convergent Gaussian integral over $\bm{\omega}$ yields
\begin{equation}
\Var_{\mathrm{MCE}}(X) = \Sigma_{XX} - \bm{\Sigma}_{X\mathcal{Q}}\,\bm{\Sigma}_{\mathcal{Q}\mathcal{Q}}^{-1}\,\bm{\Sigma}_{\mathcal{Q}X},
\label{eq:schur}
\end{equation}
where $\Sigma_{ab}$ denotes the GCE covariance matrix evaluated at $\bm{\tau}^*$. Mathematically, Eq.~\eqref{eq:schur} is the Schur complement of $\bm{\Sigma}_{\mathcal{Q}\mathcal{Q}}$ inside the joint covariance matrix $\bigl(\begin{smallmatrix} \Sigma_{XX} & \bm{\Sigma}_{X\mathcal{Q}} \\ \bm{\Sigma}_{\mathcal{Q}X} & \bm{\Sigma}_{\mathcal{Q}\mathcal{Q}} \end{smallmatrix}\bigr)$---that is, the conditional variance of a multivariate normal distribution after subtracting the linear projection of $X$ onto the conserved quantities $\hat{\bm{Q}}$. Non-Gaussian corrections from higher cumulants of $\Phi$ are suppressed by additional powers of $1/N$.

\subsection{Conservation baseline for \texorpdfstring{$\langle \Delta p_{T,i}\Delta p_{T,j} \rangle$}{<Delta pT,i Delta pT,j>}}
\label{sec:cons_baseline}

To connect Eq.~\eqref{eq:schur} to the STAR correlator, define the total transverse-momentum deviation of accepted particles in an event, $X = \sum_{k\in\mathrm{acc}} \Delta p_{T,k}$, where $\Delta p_{T,k} = p_{T,k} - \meanpT$. Using the identity $\sum_{i\neq j} \Delta p_{T,i}\Delta p_{T,j} = X^2 - \sum_k (\Delta p_{T,k})^2$ together with $\langle X \rangle = 0$ and $\langle N(N-1)\rangle \simeq N_{\mathrm{acc}}^2$, we obtain
\begin{equation*}
\correl \simeq \frac{\Var(X) - \bigl\langle \sum_k (\Delta p_{T,k})^2 \bigr\rangle}{N_{\mathrm{acc}}^2},
\end{equation*}
which differs from the event-wise normalization in Eq.~\eqref{eq:correlator} only at relative order $1/N$ [i.e., absolute order $O(1/N^2)$]. In an uncorrelated GCE Boltzmann gas, Poisson multiplicity fluctuations give $\Var_{\mathrm{GCE}}(X) = \langle \sum_k (\Delta p_{T,k})^2 \rangle$, so $\correl_{\mathrm{GCE}} = 0$ identically. Imposing global conservation via Eq.~\eqref{eq:schur} removes the fluctuating charge projections and produces a strictly nonpositive correlation,
\begin{equation}
\correl_{\mathrm{cons}} = -\frac{\bm{\Sigma}_{X\mathcal{Q}}\,\bm{\Sigma}_{\mathcal{Q}\mathcal{Q}}^{-1}\,\bm{\Sigma}_{\mathcal{Q}X}}{N_{\mathrm{acc}}^2} \le 0,
\label{eq:cons}
\end{equation}
because $\bm{\Sigma}_{\mathcal{Q}\mathcal{Q}}$ is positive definite.

Physical intuition for Eq.~\eqref{eq:cons} comes from considering a single Boltzmann species of multiplicity $N$. With energy conservation alone, Eq.~\eqref{eq:cons} gives $\correl/\meanpT^2 = -\Cov(p_T, \varepsilon)^2 / \bigl(N \meanpT^2 \langle \varepsilon^2 \rangle\bigr)$: a particle with above-average $p_T$ carries above-average energy $\varepsilon$, leaving less energy for the remaining particles and anticorrelating their $p_T$. This evaluates to $-1/(12N)$ in the massless limit and to $-(T/m)^2/(4N)$ for $T \ll m$, where rest mass dominates $\langle \varepsilon^2 \rangle \simeq m^2$. If particle (or baryon) number is conserved simultaneously with energy at $T \ll m$, rest-mass fluctuations drop out of the constrained subspace ($\langle \varepsilon^2 \rangle \to \Var(\varepsilon) = \frac{3}{2}T^2$), restoring an unsuppressed kinetic-energy anticorrelation of $-1/(6N)$. In a realistic experiment where only charged hadrons in a limited kinematic window are accepted, unaccepted neutral and forward/backward hadrons act as an energy- and charge-absorbing reservoir that dilutes the magnitude of Eq.~\eqref{eq:cons}.

\subsection{Transverse-momentum conservation}
\label{sec:pt_cons}

Global conservation of the total transverse momentum $\bm{P} = \sum_k \bm{p}_{T,k}$ over all primordial hadrons behaves differently because $\bm{P}$ is a two-dimensional vector with GCE variance $\Var(P_x) = \Var(P_y) \equiv \Sigma_P = \frac{1}{2}\sum_k \langle p_{T,k}^2 \rangle$. Conditioning on $\bm{P} = \bm{0}$ induces familiar back-to-back azimuthal correlations of order $1/N$~\cite{Borghini2000}. However, because the scalar observable $X$ depends only on $p_T$ magnitudes and is invariant under azimuthal rotations, $\Cov(X, \bm{P}) = \bm{0}$, so the $O(1/N)$ Schur complement~\eqref{eq:cons} vanishes identically.

A nonzero contribution appears only at next-to-leading order, $O(1/N^2)$. Conditioning a Poisson gas on $\bm{P} = \bm{0}$ reweights the two-particle momentum density by $\varphi(-\bm{p}_1 - \bm{p}_2)/\varphi(\bm{0})$ and each single-particle density by $\varphi(-\bm{p}_i)/\varphi(\bm{0})$, where $\varphi(\bm{P})$ is the two-dimensional Gaussian distribution of $\bm{P}$. Subtracting the disconnected product of conditioned single-particle densities yields the connected pair density
\begin{align}
\rho_1 \rho_2\, e^{-(p_1^2 + p_2^2)/2\Sigma_P} &\bigl(e^{-\bm{p}_1\cdot\bm{p}_2/\Sigma_P} - 1\bigr) \nonumber\\
&\simeq \rho_1 \rho_2 \left[-\frac{\bm{p}_1\cdot\bm{p}_2}{\Sigma_P} + \frac{(\bm{p}_1\cdot\bm{p}_2)^2}{2\Sigma_P^2}\right].
\label{eq:pt_pair}
\end{align}
Upon integrating over particle azimuths, the dipole recoil term $-\bm{p}_1\cdot\bm{p}_2/\Sigma_P \propto \cos(\phi_1 - \phi_2)$ averages to zero, whereas the quadrupole recoil term $(\bm{p}_1\cdot\bm{p}_2)^2/(2\Sigma_P^2) \propto \cos^2(\phi_1 - \phi_2)$ survives with azimuthal average $1/2$: intuitively, if one particle has an unusually large $p_T$, recoil forces other particles to carry larger $p_T$ magnitudes in the opposite hemisphere. Integrating against $\Delta p_{T,1}\Delta p_{T,2}/N_{\mathrm{acc}}^2$ gives
\begin{equation}
\correl_{P_T} = \frac{G_P^2}{4\Sigma_P^2 N_{\mathrm{acc}}^2}, \quad G_P = \Cov\bigl(X, {\textstyle\sum_k} p_{T,k}^2\bigr),
\label{eq:pt_cons}
\end{equation}
which reduces for a single undecayed species to $\Cov(p_T, p_T^2)^2 / \bigl(N^2 \langle p_T^2 \rangle^2\bigr) > 0$. Non-Gaussian fourth-cumulant corrections to $\varphi(\bm{P})$ enter the azimuthally symmetric pair correlation only at $O(1/N^3)$.

\subsection{Fixed species multiplicities and the mixing term}
\label{sec:mixing}

\begin{figure*}[t]
\centering
\includegraphics[width=\textwidth]{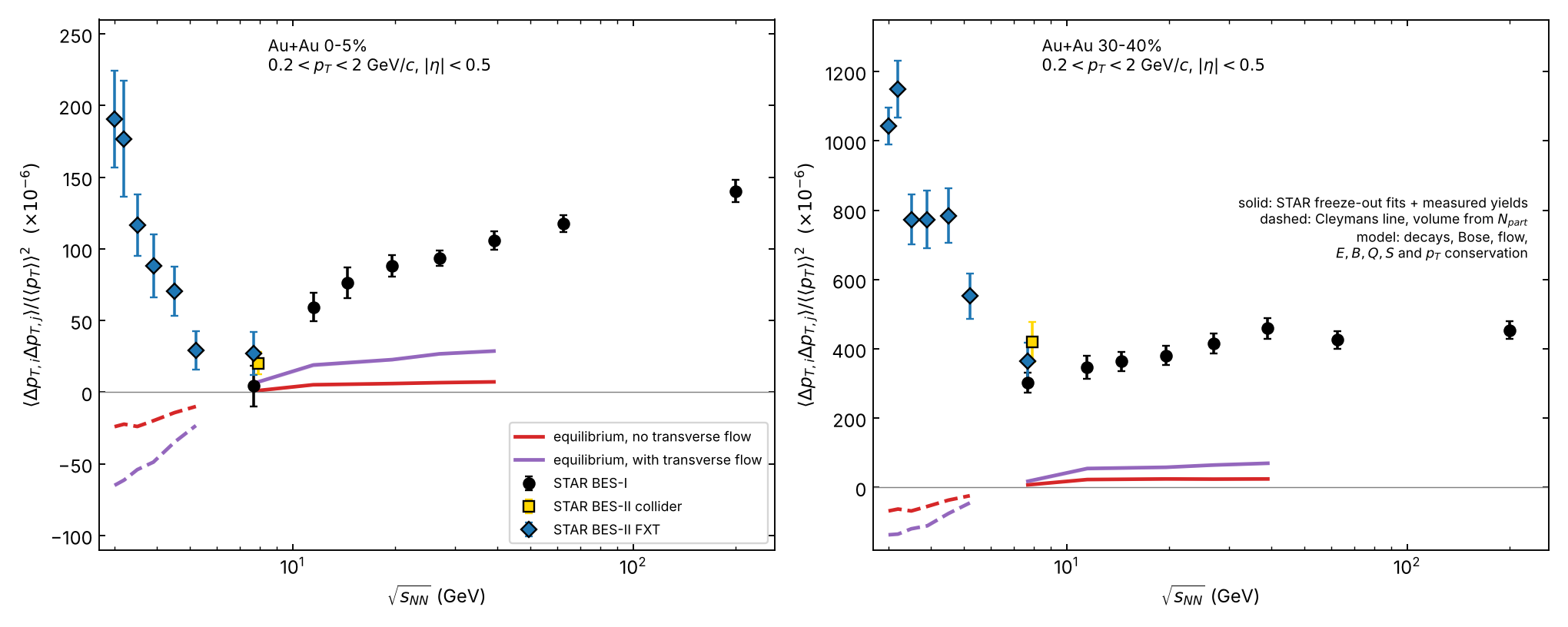}
\caption{Scaled transverse-momentum correlator $\correl/\meanpT^2$ in 0--5\% (left) and 30--40\% (right) Au+Au collisions as a function of collision energy $\sNN$. Symbols show STAR measurements from BES-I~\cite{STAR2019} and BES-II collider and FXT runs~\cite{STAR2026} ($C_{p_T}^2$, with statistical and systematic uncertainties added in quadrature). Curves show the microcanonical equilibrium baseline without (red) and with (purple) transverse radial flow: solid segments ($\sNN = 7.7$--$39$~GeV) employ STAR freeze-out fits and measured midrapidity yields, whereas dashed segments ($\sNN = 3$--$5.2$~GeV) use the parameterized freeze-out curve, $\langle N_{\mathrm{part}} \rangle$, and fixed $\beta_s$ (the two segments are left disconnected between $5.2$ and $7.7$~GeV due to the change in input prescriptions). Shaded bands indicate $\pm 2\sigma$ Monte Carlo statistical uncertainties.}
\label{fig:main}
\end{figure*}

Our Schur-complement formalism also clarifies how multi-species ``mixing'' behaves in and out of equilibrium. Suppose the accepted multiplicity $N_s$ of every hadron species $s$ (pions, kaons, protons) were artificially held fixed from event to event. Treating the $N_s$ as conserved quantities in Eq.~\eqref{eq:schur}, with $\Sigma_{X N_s} = N_s(\langle p_T \rangle_s - \meanpT)$ and $\Sigma_{N_s N_{s'}} = \delta_{ss'} N_s$, immediately yields
\begin{equation}
\correl_{\mathrm{fix}} = -\frac{1}{N_{\mathrm{acc}}} \sum_s x_s \bigl(\langle p_T \rangle_s - \meanpT\bigr)^2,
\label{eq:fix}
\end{equation}
where $x_s = N_s / N_{\mathrm{acc}}$. Up to $N \to N - 1$, Eq.~\eqref{eq:fix} matches the direct evaluation of Eq.~\eqref{eq:correlator} at fixed composition; for a binary proton--pion system it reduces to $-x_p x_\pi (\langle p_T \rangle_p - \langle p_T \rangle_\pi)^2 / N$, which attains its maximum magnitude at $x_p = x_\pi = 1/2$ and corresponds to the mixing mechanism proposed in Ref.~\cite{Reichert2026}.

In a grand-canonical thermal gas, however, species multiplicities are not fixed: independent Poisson fluctuations of $N_s$ contribute an identical positive variance $+\sum_s N_s(\langle p_T \rangle_s - \meanpT)^2$ to $\Var_{\mathrm{GCE}}(X)$, canceling Eq.~\eqref{eq:fix} identically so that any Poisson mixture satisfies $\correl_{\mathrm{GCE}} = 0$. In a true microcanonical equilibrium ensemble, hadronic composition is constrained only indirectly through the conserved charges $(E,B,Q,S)$: net baryon number fixes the sum over \emph{all} baryons minus antibaryons across full phase space (including unmeasured neutrons, hyperons, and excited resonances) and places no direct constraint on meson numbers. Consequently, the physical equilibrium counterpart of species mixing is already contained within the Schur projection of Eq.~\eqref{eq:cons} and, as quantified in Sec.~\ref{sec:results}, is an order of magnitude weaker than Eq.~\eqref{eq:fix}.

\section{Model}
\label{sec:model}

\subsection{Hadron spectrum and resonance decays}
\label{sec:spectrum}

\begin{table*}[t]
\caption{Equilibrium model inputs and results for 0--5\% central Au+Au collisions: midrapidity volume $V_0 = dV/d\eta_s|_{\eta_s=0}$ ($\mathrm{fm}^3$), fluid-rapidity Gaussian width $\sigma$, and mean accepted charged multiplicity $N_{\mathrm{acc}}$. Model predictions (without and with transverse flow $\beta_s$) and STAR measurements ($C_{p_T}^2$) for $\correl/\meanpT^2$ are listed in units of $10^{-6}$; the $7.7$~GeV datum is from BES-II (collider). The final column gives the relative surface-velocity fluctuation $\sigma_\beta/\beta_s$ at fixed fireball energy needed to bridge the gap between the flowing equilibrium baseline and the data. Monte Carlo statistical uncertainties on the model values are $0.2$--$1.5$.}
\label{tab:central}
\begin{ruledtabular}
\begin{tabular}{rrrrrrrrr}
$\sNN$ (GeV) & $V_0$ ($\mathrm{fm}^3$) & $\sigma$ & $N_{\mathrm{acc}}$ & No flow & $\beta_s$ & Flow & STAR & $\sigma_\beta/\beta_s$ (\%) \\
\colrule
3    & 5725 & 0.43 &  60 & $-24.0$ & 0.45 & $-65.0$ & $190 \pm 34$ & 10 \\
3.2  & 4735 & 0.46 &  65 & $-22.3$ & 0.45 & $-61.3$ & $177 \pm 40$ & 10 \\
3.5  & 3740 & 0.50 &  72 & $-23.9$ & 0.45 & $-54.0$ & $117 \pm 21$ &  9 \\
3.9  & 2917 & 0.55 &  80 & $-19.9$ & 0.45 & $-48.8$ &  $88 \pm 22$ &  9 \\
4.5  & 2216 & 0.62 &  93 & $-14.2$ & 0.45 & $-34.8$ &  $71 \pm 17$ &  9 \\
5.2  & 1772 & 0.68 & 106 &  $-9.9$ & 0.45 & $-23.3$ &  $29 \pm 13$ &  8 \\
\colrule
7.7  & 1170 & 0.80 & 162 &  $+0.9$ & 0.69 &  $+6.3$ &  $20 \pm 8$  &  4 \\
11.5 & 1221 & 1.08 & 205 &  $+5.2$ & 0.70 & $+19.0$ &  $59 \pm 10$ &  6 \\
19.6 & 1152 & 1.47 & 259 &  $+6.1$ & 0.69 & $+22.8$ &  $88 \pm 8$  &  8 \\
27   & 1116 & 1.91 & 278 &  $+6.7$ & 0.72 & $+26.8$ &  $94 \pm 5$  &  9 \\
39   & 1178 & 2.79 & 292 &  $+7.2$ & 0.74 & $+28.7$ & $106 \pm 6$  &  9 \\
\end{tabular}
\end{ruledtabular}
\end{table*}

We evaluate the equilibrium baseline using the PDG2025 particle list and decay table implemented in \textsc{Thermal-FIST}~\cite{Vovchenko2019}, comprising 248 hadron states plus their antiparticles. Primordial hadrons are sampled by Monte Carlo and decayed sequentially---using exact two-body kinematics and phase-space weights for three- or more-body channels---down to strong- and electromagnetic-interaction stable states (weak decays are excluded). Branching fractions are renormalized to the channels kinematically open at each resonance's pole mass, while Breit--Wigner mass sampling is performed as a systematic check.

Resonance decays affect $\correl$ in two distinct ways. First, daughters originating from the same parent resonance are kinematically correlated within the same event, adding a direct two-particle decay term
\begin{equation*}
V \sum_i \int \frac{d^3p}{(2\pi)^3}\, g_i f_i \Bigl\langle {\textstyle\sum_{d\neq d'}} \Delta p_{T,d}\Delta p_{T,d'} \Bigr\rangle
\end{equation*}
to the numerator of $\correl$. Second, in the conservation matrix of Eq.~\eqref{eq:cons}, each primordial hadron enters $\bm{\Sigma}_{X\mathcal{Q}}$ through the decay-averaged accepted $p_T$ deviation of its final-state daughters, whereas $\bm{\Sigma}_{\mathcal{Q}\mathcal{Q}}$ receives contributions from all primordial hadrons in the fireball regardless of acceptance.

\subsection{Bose-Einstein statistics}
\label{sec:bose}

Quantum statistics are included for light primordial mesons ($m < 0.9$~GeV) by replacing the Boltzmann distribution with the Bose-Einstein occupation number $f = 1/(e^{(E-\mu)/T} - 1)$. Because occupation fluctuations of bosonic modes obey $\Cov(n_{\bm{p}}, n_{\bm{p}'}) = \delta_{\bm{p}\bm{p}'} f(1+f)$, all primordial meson entries in the GCE covariance matrices acquire Bose enhancement weights $f(1+f)$. In addition, subtracting the self-correlation sum $\langle \sum_k (\Delta p_{T,k})^2 \rangle \propto f$ from $\Var(X) \propto f(1+f)$ leaves a positive two-particle Wick contraction proportional to $f^2$ among accepted stable bosons,
\begin{align}
\correl_{\mathrm{Bose}} &= \frac{V}{N_{\mathrm{acc}}^2} \sum_{i\in\{\pi^\pm, K^\pm\}} g_i \nonumber\\
&\quad \times \int \frac{d^3p}{(2\pi)^3}\, f_i^2 \bigl(p_T - \meanpT\bigr)^2 > 0.
\label{eq:bose}
\end{align}
We evaluate Eq.~\eqref{eq:bose} in the local rest frame of each fluid element, which is justified whenever the thermal wavelength is small compared with the homogeneity length of the fireball. Quantum-statistical correlations of short-lived resonances and exchange terms between decay daughters of distinct parents are neglected.

\subsection{Collective expansion and kinematic acceptance}
\label{sec:flow}

\begin{figure*}[t]
\centering
\includegraphics[width=\textwidth]{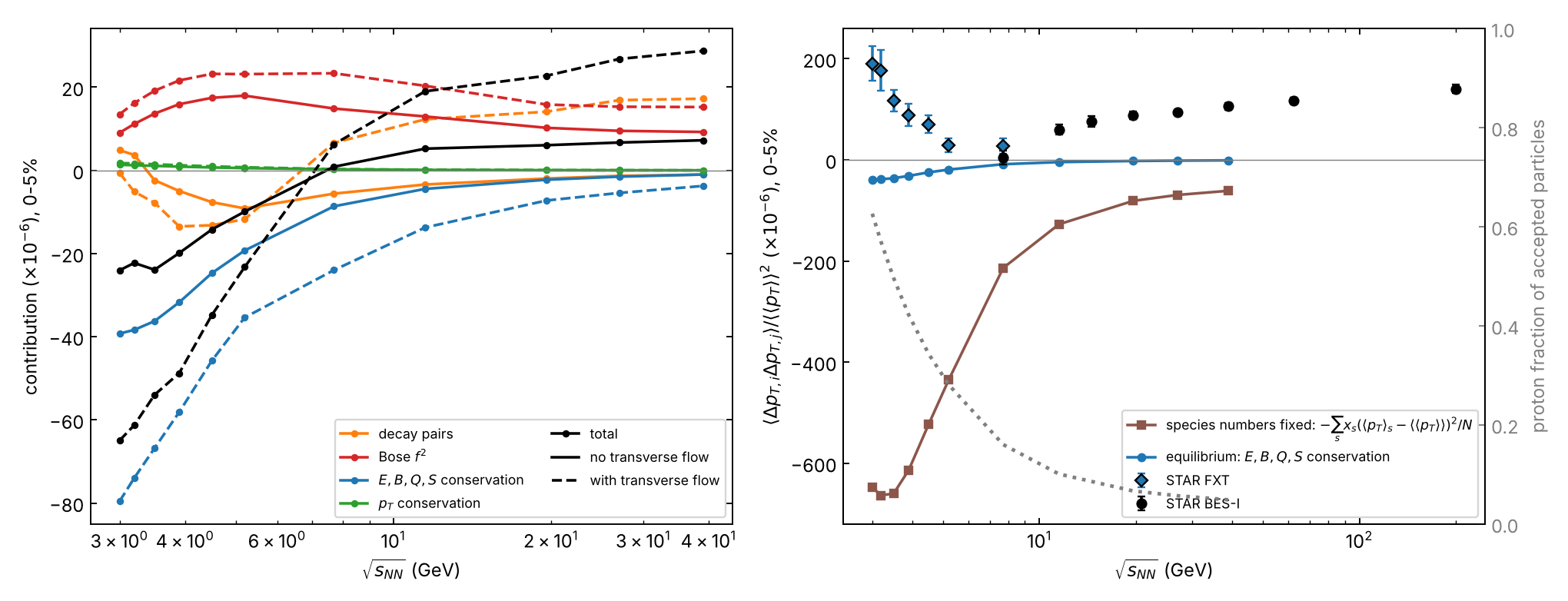}
\caption{Left: anatomy of the equilibrium baseline for $\correl/\meanpT^2$ in 0--5\% central collisions without (solid) and with (dashed) transverse radial flow, showing separate contributions from resonance-decay pairs, Bose-Einstein statistics, $(E,B,Q,S)$ conservation, and $p_T$ conservation. Right: comparison between the fixed-multiplicity mixing formula of Eq.~\eqref{eq:fix} (brown squares), the self-consistent equilibrium $(E,B,Q,S)$ conservation term without flow (blue circles), and STAR data; the dotted curve (right axis) tracks the accepted proton fraction $x_p$.}
\label{fig:decomp}
\end{figure*}

\begin{table*}[t]
\caption{Same as Table~\ref{tab:central} for 30--40\% centrality. Monte Carlo statistical uncertainties on the model values are $1$--$5$.}
\label{tab:midcentral}
\begin{ruledtabular}
\begin{tabular}{rrrrrrrrr}
$\sNN$ (GeV) & $V_0$ ($\mathrm{fm}^3$) & $\sigma$ & $N_{\mathrm{acc}}$ & No flow & $\beta_s$ & Flow & STAR & $\sigma_\beta/\beta_s$ (\%) \\
\colrule
3    & 1379 & 0.57 & 16 & $-67.7$ & 0.35 & $-136.5$ & $1043 \pm 53$ & 31 \\
3.2  & 1140 & 0.62 & 17 & $-62.1$ & 0.35 & $-134.4$ & $1149 \pm 81$ & 35 \\
3.5  &  901 & 0.68 & 19 & $-67.7$ & 0.35 & $-119.1$ &  $773 \pm 72$ & 31 \\
3.9  &  703 & 0.74 & 21 & $-54.6$ & 0.35 & $-110.5$ &  $773 \pm 83$ & 35 \\
4.5  &  534 & 0.83 & 25 & $-36.4$ & 0.35 &  $-75.0$ &  $784 \pm 78$ & 40 \\
5.2  &  427 & 0.92 & 28 & $-23.7$ & 0.35 &  $-44.4$ &  $552 \pm 66$ & 38 \\
\colrule
7.7  &  308 & 1.08 & 47 &  $+7.7$ & 0.52 &  $+18.0$ &  $420 \pm 57$ & 23 \\
11.5 &  253 & 1.44 & 57 & $+23.0$ & 0.54 &  $+54.9$ &  $346 \pm 33$ & 22 \\
19.6 &  279 & 1.98 & 73 & $+24.7$ & 0.54 &  $+58.4$ &  $380 \pm 28$ & 22 \\
27   &  266 & 2.50 & 79 & $+24.2$ & 0.57 &  $+64.8$ &  $416 \pm 29$ & 23 \\
39   &  283 & 3.42 & 84 & $+24.7$ & 0.59 &  $+69.9$ &  $459 \pm 30$ & 23 \\
\end{tabular}
\end{ruledtabular}
\end{table*}

To assess the role of collective expansion, particle momenta are sampled at the chemical freeze-out temperature $T_{\mathrm{ch}}$ in the local rest frame of each fluid element and boosted both transversely and longitudinally. Transversely, we adopt a linear blast-wave velocity profile $v(r) = \beta_s r/R$ with volume-emission weight $2r\,dr/R^2$, corresponding to a mean radial velocity $\langle \beta \rangle = \frac{2}{3}\beta_s$. For $\sNN = 7.7$--$39$~GeV, we set $\beta_s = 1.5\langle \beta \rangle$ using $\langle \beta \rangle$ from STAR kinetic freeze-out fits~\cite{STAR2017}; at FXT energies, where tabulated blast-wave fits are not yet available, we take $\beta_s = 0.45$ in 0--5\% and $0.35$ in 30--40\% collisions. Combining these flow velocities with $T_{\mathrm{ch}} > T_{\mathrm{kin}}$ yields $\meanpT \simeq 0.60$~GeV/$c$ in central collisions at BES-I energies, providing a generous upper-limit estimate of radial-flow effects.

Longitudinally, fluid elements boost with their space-time rapidity $\eta_s$ (Bjorken scaling) weighted by a Gaussian volume profile $dV/d\eta_s = V_0 \exp(-\eta_s^2 / 2\sigma^2)$. A final-state charged hadron with local rest-frame rapidity $y^*$ is accepted if its boosted laboratory momentum falls inside the STAR fiducial window ($|\eta| < 0.5$, $0.2 < p_T < 2.0$~GeV/$c$), defining a Gaussian-weighted effective acceptance length in $\eta_s$ for both single hadrons and correlated decay pairs. Meanwhile, the full fireball volume $\int d\eta_s\,(dV/d\eta_s) = \sqrt{2\pi}\sigma V_0$ governs the global charge susceptibilities in $\bm{\Sigma}_{\mathcal{Q}\mathcal{Q}}$.

\subsection{Freeze-out inputs}
\label{sec:inputs}

At collider energies ($\sNN = 7.7$--$39$~GeV), we take the STAR chemical freeze-out parameters $(T, \mu_B, \mu_S, \gamma_S)$, midrapidity $\pi^\pm, K^\pm, p, \bar{p}$ yields, and $\langle N_{\mathrm{part}} \rangle$ for 0--5\% and 30--40\% centrality directly from Ref.~\cite{STAR2017}. The two geometric parameters $V_0$ and $\sigma$ are fixed uniquely at each energy by requiring that (i)~the post-decay model reproduces the summed midrapidity $dN/dy|_{y=0}$ of $\pi^\pm, K^\pm, p, \bar{p}$, and (ii)~the total net baryon number in $\sqrt{2\pi}\sigma V_0$ equals $\langle N_{\mathrm{part}} \rangle$. In 0--5\% collisions, the resulting ratio $\sigma/y_{\mathrm{beam}}$ grows smoothly from $0.38$ at $7.7$~GeV to $0.75$ at $39$~GeV.

At FXT energies ($\sNN = 3$--$5.2$~GeV), where numerical tables of identified-hadron yields and freeze-out fits are not yet publicly available~\cite{Labonte2026}, we adopt the chemical freeze-out parametrization of Cleymans \emph{et al.}~\cite{Cleymans2006} with $\gamma_S = 1$, fixing $N_{\mathrm{part}} = 340$ ($110$) and $\sigma/y_{\mathrm{beam}} = 0.41$ ($0.55$) for 0--5\% (30--40\%) collisions---matching the average $\sigma/y_{\mathrm{beam}}$ extracted at $7.7$ and $11.5$~GeV. Tested at $7.7$~GeV, this prescription reproduces the fitted $V_0$ to within 5\%. Because every $O(1/N)$ baseline term scales inversely with $V_0$, the FXT baseline carries an overall normalization uncertainty of roughly 25\%, which leaves all signs and qualitative conclusions unchanged.

\section{Results}
\label{sec:results}

\subsection{Comparison with STAR data}
\label{sec:comparison}

Figure~\ref{fig:main} and Tables~\ref{tab:central} and~\ref{tab:midcentral} compare our equilibrium baseline for $\correl/\meanpT^2$ with the STAR BES-I~\cite{STAR2019} and BES-II~\cite{STAR2026} measurements ($C_{p_T}^2$). In 0--5\% central collisions across $\sNN = 11.5$--$39$~GeV, the static equilibrium correlator reaches only $(5\text{--}7) \times 10^{-6}$. Adding strong radial flow boosts the baseline to $(19\text{--}29) \times 10^{-6}$, which nevertheless remains a factor of $3.1$--$3.9$ below the experimental data. Near the minimum of the data at $\sNN = 7.7$~GeV, the flowing baseline ($6.3 \times 10^{-6}$) lies below the precise BES-II collider point ($(20 \pm 8) \times 10^{-6}$) while remaining consistent within errors with the earlier BES-I measurement.

The discrepancy becomes dramatic in the FXT regime ($\sNN = 3$--$5.2$~GeV): whereas the measured correlator turns sharply upward, reaching $(190 \pm 34) \times 10^{-6}$ at $3$~GeV, the equilibrium baseline is \emph{negative} everywhere below $7$~GeV---ranging from $-(10\text{--}24) \times 10^{-6}$ without flow to $-(23\text{--}65) \times 10^{-6}$ with radial flow. The same qualitative picture holds in 30--40\% midcentral collisions (Table~\ref{tab:midcentral}), where the flowing baseline yields $(55\text{--}70) \times 10^{-6}$ against $(346\text{--}459) \times 10^{-6}$ in the data at $\sNN = 11.5$--$39$~GeV (a factor of $\approx 6.5$ shortfall), and again becomes negative across FXT energies where the data climb above $10^{-3}$.

\subsection{Decomposition of equilibrium effects}
\label{sec:decomp}

Figure~\ref{fig:decomp} (left) and Table~\ref{tab:decomp} break the 0--5\% baseline down into its four physical mechanisms:
\begin{enumerate}
\item \emph{Global $(E,B,Q,S)$ conservation} is negative at all energies, as required by Eq.~\eqref{eq:cons}. Its magnitude peaks at low $\sNN$, where $N_{\mathrm{acc}}$ is smallest and massive nucleons dominate the accepted sample, and is amplified by radial flow because collective boosting strengthens $\Cov(p_T, \varepsilon)$.
\item \emph{Resonance-decay pairs} yield a modest negative correlation in a static gas---since two-body decay kinematics at low parent momentum tend to anticorrelate daughter $p_T$ magnitudes---but turn positive at $\sNN \ge 7.7$~GeV once radial flow is included, because sibling daughters inherit the common radial boost of their parent resonance.
\item \emph{Bose-Einstein statistics} generate a positive correlation of $(9\text{--}23) \times 10^{-6}$ in central collisions via Eq.~\eqref{eq:bose}. Notably, in a static gas without resonance feed-down this Bose signal would be roughly three times larger; because most final-state pions originate from resonance decays rather than primordial emission, the $1/N_{\mathrm{acc}}^2$ dilution suppresses the primordial Wick term substantially.
\item \emph{Transverse-momentum conservation} [Eq.~\eqref{eq:pt_cons}] is positive as expected at $O(1/N^2)$ but numerically negligible everywhere: $(0.01\text{--}1.8) \times 10^{-6}$ in 0--5\% collisions, reaching at most $16 \times 10^{-6}$ in the lowest-multiplicity 30--40\% FXT bin ($N_{\mathrm{acc}} = 16$).
\end{enumerate}

\begin{table}[!htbp]
\caption{Decomposition of $\correl/\meanpT^2$ ($\times 10^{-6}$) in 0--5\% collisions with transverse flow as a function of $\sNN$ (GeV): resonance-decay pairs, Bose-Einstein Wick term, $(E,B,Q,S)$ conservation, $p_T$ conservation, and their sum. The final two columns list the accepted proton fraction $x_p$ and the fixed-multiplicity mixing term of Eq.~\eqref{eq:fix} (evaluated without flow).}
\label{tab:decomp}
\begin{ruledtabular}
\begin{tabular}{rrrrrrrr}
$\sNN$ & Pairs & Bose & $E,B,Q,S$ & $p_T$ & Total & $x_p$ & Fixed $N_s$ \\
\colrule
3    &  $-0.6$ & $+13.5$ & $-79.6$ & $+1.77$ & $-65.0$ & 0.63 & $-648$ \\
3.5  &  $-7.8$ & $+19.2$ & $-66.8$ & $+1.47$ & $-54.0$ & 0.50 & $-660$ \\
4.5  & $-13.2$ & $+23.2$ & $-45.8$ & $+0.97$ & $-34.8$ & 0.34 & $-523$ \\
5.2  & $-11.7$ & $+23.1$ & $-35.4$ & $+0.74$ & $-23.3$ & 0.28 & $-435$ \\
7.7  &  $+6.6$ & $+23.3$ & $-24.0$ & $+0.30$ &  $+6.3$ & 0.16 & $-214$ \\
11.5 & $+12.3$ & $+20.3$ & $-13.7$ & $+0.15$ & $+19.0$ & 0.10 & $-127$ \\
19.6 & $+14.1$ & $+15.8$ &  $-7.2$ & $+0.06$ & $+22.8$ & 0.07 &  $-81$ \\
39   & $+17.2$ & $+15.2$ &  $-3.7$ & $+0.02$ & $+28.7$ & 0.05 &  $-61$ \\
\end{tabular}
\end{ruledtabular}
\end{table}
The sign change of the total baseline near $\sNN \approx 7$~GeV thus reflects a competition between positive Bose and flow-boosted decay correlations at collider energies and strong negative $(E,B,Q,S)$ conservation at FXT energies.

\subsection{Baryon--meson mixing in equilibrium}
\label{sec:baryon_meson}

Across our energy scan, the accepted charged-hadron sample evolves smoothly from proton-dominated matter ($x_p \approx 0.63$ at $3$~GeV, passing through equal proton and meson shares near $3.5$~GeV) to pion-dominated matter ($x_p \approx 0.05$ at $39$~GeV). As shown in Fig.~\ref{fig:decomp} (right) and Table~\ref{tab:decomp}, freezing the accepted multiplicities of individual species event by event via Eq.~\eqref{eq:fix} would produce a massive negative correlation: $-(648\text{--}660) \times 10^{-6}$ at $3$--$3.5$~GeV, relaxing to $-61 \times 10^{-6}$ at $39$~GeV.

When species multiplicities are allowed to fluctuate in thermal and chemical equilibrium subject only to global $(E,B,Q,S)$ conservation, the resulting correlation without flow is just $-(1\text{--}39) \times 10^{-6}$---smaller by a factor of $15$--$60$. Thus, while an equilibrated HRG naturally incorporates the baryon-to-meson crossover, equilibrium multiplicity fluctuations erase nearly all of the fixed-species mixing anticorrelation, and the small surviving conservation effect has the wrong sign to explain the FXT rise.

\subsection{Centrality scaling and radial-flow fluctuations}
\label{sec:flow_fluct}

Because every equilibrium baseline contribution scales as $1/N$ [or $1/N^2$ for Eq.~\eqref{eq:pt_cons}], going from 0--5\% to 30--40\% centrality---where $N_{\mathrm{acc}}$ drops by a factor of $3.5$ across $\sNN = 11.5$--$39$~GeV---increases the model prediction by a factor of $3.4$--$4.4$ without flow and $2.4$--$2.9$ with flow (Tables~\ref{tab:central} and~\ref{tab:midcentral}). Over the same energy range, the STAR data grow by a factor of $4.3$--$5.9$, exceeding pure $1/N$ scaling.

To estimate the magnitude of dynamical collective fluctuations required by the data, suppose the surface velocity $\beta_s$ fluctuates event by event around its mean. By the law of total covariance, parameter fluctuations add an $N$-independent positive contribution,
\begin{equation*}
\frac{\Var(\langle p_T \rangle_\theta)}{\meanpT^2} = \left(\frac{\partial \ln\meanpT}{\partial \ln\beta_s}\right)^{\!2} \left(\frac{\sigma_\beta}{\beta_s}\right)^{\!2}.
\end{equation*}
Crucially, global energy conservation confines event-by-event velocity fluctuations to trajectories of constant total fireball energy, along which a faster collective expansion must be compensated by a lower freeze-out temperature. Evaluating this constrained derivative within our full HRG model gives $\partial \ln\meanpT / \partial \ln\beta_s|_E \approx 0.10$ in 0--5\% and $0.08$ in 30--40\% collisions at collider energies---far weaker than the unconstrained fixed-temperature sensitivities of $0.47$--$0.53$ and $0.29$--$0.33$, respectively.

As listed in the last columns of Tables~\ref{tab:central} and~\ref{tab:midcentral}, matching the excess of the STAR data over our flowing equilibrium baseline requires $\sigma_\beta/\beta_s \approx 6$--$10\%$ in 0--5\% central collisions (dipping to $4\%$ at $7.7$~GeV) and $22$--$23\%$ in 30--40\% collisions across $\sNN = 7.7$--$39$~GeV, rising to $31$--$40\%$ in midcentral FXT collisions. At $\sNN = 11.5$--$39$~GeV, the ratio of required relative velocity fluctuations between 30--40\% and 0--5\% collisions is $2.6$--$3.4$, substantially steeper than the independent-source expectation $\sqrt{N_{\mathrm{part}}(\text{0--5\%})/N_{\mathrm{part}}(\text{30--40\%})} \approx 1.8$. Neglecting the constant-energy constraint would underestimate the required $\sigma_\beta/\beta_s$ by a factor of $2.5$--$5$.

\subsection{Model sensitivity and approximations}
\label{sec:sensitivity}

We verified that reasonable variations of the model ingredients do not alter our conclusions. Sampling finite Breit--Wigner widths for primordial resonances lowers the flowing baseline by $20\%$ at $19.6$~GeV ($22.8 \to 18.1 \times 10^{-6}$) and makes the $3$~GeV value even more negative ($-65 \to -83 \times 10^{-6}$). Varying $\sigma/y_{\mathrm{beam}}$ by $\pm 25\%$ at FXT energies rescales the baseline by $\mp 25\%$ without changing its sign. Purely global volume fluctuations at fixed intensive freeze-out parameters leave $\correl$ unchanged, because particle momenta remain uncorrelated at any fixed volume.

Several simplifying approximations merit note: decay daughters are evaluated at pole masses; local momentum spectra employ $T_{\mathrm{ch}}$ with a uniform volume-emission blast-wave profile; Bose-Einstein statistics are restricted to primordial light mesons; light nuclei, which carry a substantial fraction of the baryon number at $\sNN = 3$~GeV, are omitted; strangeness is treated grand-canonically prior to global projection despite canonical strangeness suppression at $3$~GeV~\cite{STAR2022}; and higher-order non-Gaussian saddle-point terms as well as final-state Coulomb interactions are neglected. At FXT energies, freeze-out parameters and fireball volumes rely on smooth parametrizations pending final identified-particle spectra.

\section{Summary}
\label{sec:summary}

Using an inverse Laplace and Fourier representation of the microcanonical ensemble, we derived the leading $O(1/N)$ effect of global conservation laws on momentum-weighted observables as a Schur complement of grand-canonical covariances. Applied to the STAR two-particle transverse-momentum correlator $\correl$, global conservation of energy, baryon number, electric charge, and strangeness yields a strictly negative $O(1/N)$ contribution, whereas global transverse-momentum conservation vanishes at $O(1/N)$ by azimuthal symmetry and enters only as a tiny positive $O(1/N^2)$ recoil effect. While artificially fixing species multiplicities event by event generates a large negative baryon--meson mixing term, thermal multiplicity fluctuations in equilibrium cancel that variance, leaving only the much weaker microcanonical charge-conservation projection.

Evaluating the complete equilibrium baseline---combining PDG2025 resonance decays, Bose-Einstein statistics, longitudinal Bjorken and transverse radial flow, and global conservation laws---across $\sNN = 3$--$39$~GeV demonstrates that an equilibrated hadron gas at fixed intensive freeze-out parameters cannot explain the STAR measurements. Even with strong radial flow, the equilibrium correlator falls short of the data by a factor of $3$--$4$ in 0--5\% collisions and $\approx 6.5$ in 30--40\% collisions at $\sNN \ge 11.5$~GeV, and turns negative across $\sNN = 3$--$5.2$~GeV where the measured FXT correlations surge strongly positive. Attributing the excess correlation to event-by-event radial-flow fluctuations along trajectories of constant total energy requires $\sigma_\beta/\beta_s \approx 6$--$10\%$ in central and $\approx 22\%$ in midcentral collider-energy collisions, exhibiting a centrality growth stronger than independent-source scaling. Subtracting this quantitative equilibrium baseline---particularly its negative conservation offset in the high-baryon-density FXT regime---is essential before interpreting transverse-momentum correlations in terms of equation-of-state softening or QCD critical fluctuations.

\section*{Data Availability}
The experimental data analyzed in this work are publicly available from Refs.~\cite{STAR2017,STAR2019,STAR2026}. All numerical baseline results supporting the findings of this article are tabulated in Tables~\ref{tab:central}--\ref{tab:decomp}; simulation code and additional outputs are available from the corresponding author upon reasonable request.

\section*{AI Use Disclosure}
During the preparation of this manuscript, the author used Google Gemini (via Google Antigravity) and Anthropic Claude to assist with analytical formula verification, bibliographic cross-checking, and manuscript editing for clarity. All AI-assisted output was directed, reviewed, and independently verified by the author, who takes full responsibility for the accuracy and integrity of all calculations, figures, and text.

\bibliography{refs}

\end{document}